\documentclass[12pt]{article}
\usepackage{epsf,epsfig,psfrag,graphicx}
\usepackage[figuresright]{rotating}
\usepackage{amsmath}

\newcommand{\be}{\begin{equation}}
\newcommand{\ee}{\end{equation}}
\newcommand{\bea}{\begin{eqnarray}}
\newcommand{\eea}{\end{eqnarray}}
\def\aprle{\buildrel < \over {_{\sim}}}
\def\aprge{\buildrel > \over {_{\sim}}}
\begin{document}  
\topmargin 0pt
\oddsidemargin=-0.4truecm
\evensidemargin=-0.4truecm
\renewcommand{\thefootnote}{\fnsymbol{footnote}}
\newpage
\setcounter{page}{0}
\begin{titlepage}   
\begin{flushright}
hep-ph/0412029
\end{flushright}
\vspace*{2cm}

\begin{center}      
{\Large \bf
Three-flavour effects and CP- and T-violation 
\vspace*{0.16cm} in
neutrino oscillations 
}
\vspace{0.8cm}

{\large
E. Kh. Akhmedov\footnote{On leave from National Research Centre Kurchatov 
Institute, Moscow 123182, Russia. Present address: ICTP, Strada 
Costiera 11, 34014 Trieste, Italy. E-mail: akhmedov@ictp.trieste.it}} \\
\vspace{0.15cm}   
{\em AHEP Group, Instituto de F\'{\i}sica Corpuscular -- C.S.I.C. \\
Universitat de Val{\`e}ncia, Edificio Institutos de Paterna \\
 Apt 22085, E--46071 Val{\`e}ncia, Spain}
\end{center}
\vglue 0.8truecm

\begin{abstract}
Some theoretical aspects of 3-flavour (3f) neutrino oscillations are reviewed. 
These include: 
general properties of 3f oscillation probabilities; matter effects in 
$\nu_\mu \leftrightarrow \nu_\tau$ oscillations; 3f 
effects in oscillations of solar, atmospheric, reactor and supernova 
neutrinos and in accelerator long-baseline experiments; CP and T 
violation in neutrino oscillations in vacuum and in matter, and the 
problem of $U_{e3}$. 
\end{abstract}
\vspace{.5cm}
\centerline{Pacs numbers: 14.60.Pq, 14.60.Lm, 26.65+t}
\end{titlepage}
\renewcommand{\thefootnote}{\arabic{footnote}}
\setcounter{footnote}{0}
\section{Introduction}

Explanation of the solar, atmospheric, reactor and accelerator neutrino data 
\cite{Concha} (with the exception of the LSND result, which still awaits its 
confirmation) 
in terms of neutrino oscillations requires at least three neutrino species, 
and in fact three neutrino species are known to exist -- $\nu_e$, $\nu_\mu$ 
and $\nu_\tau$. Yet, until a few years ago most studies of neutrino 
oscillations were performed in the 2-flavour framework. There were essentially 
two reasons for that: (i) simplicity -- there are much fewer parameters 
in the 2-flavour case than in the 3-flavour one, and  the expressions for the
transition probabilities are much simpler and by far more tractable, and 
(ii) the hierarchy $\Delta m_{\rm sol} \ll \Delta m_{\rm atm}$ and the 
smallness of the leptonic mixing parameter $|U_{e3}|$, which allow to 
effectively decouple different oscillation channels. 
The 2-flavour approach indeed proved to be a good first approximation, 
which is a consequence of the above point (ii).  

However, the increased accuracy of the available and especially of expected
neutrino data makes it very important to take  even relatively small effects 
in neutrino oscillations into account. As we shall see, 3-flavour (3f) 
effects can lead to corrections up to $\sim 10\%$ to 2-flavour oscillation 
probabilities, which is comparable to the accuracy of the present-day 
neutrino data. 
In addition, effects specific to $\ge 3$ flavour neutrino oscillations, such 
as CP and T violation, are of great interest and being widely discussed 
now. All this makes 3f analyses of neutrino oscillations mandatory. 

In my talk some theoretical issues pertaining to 3f neutrino oscillations 
are reviewed. The topics that are discussed include: 
general properties of 3f oscillation probabilities; 
matter effects in $\nu_\mu \leftrightarrow \nu_\tau$ 
oscillations; 3f effects in oscillations of solar, atmospheric, reactor 
and supernova neutrinos and in accelerator long-baseline experiments; 
CP and T violation in neutrino oscillations in vacuum and in matter;
the problem of $U_{e3}$.  

\section{Leptonic mixing and neutrino oscillations}
The leptonic mixing matrix $U$ connects neutrino flavour eigenstates 
$|\nu_a\rangle$ ($a=e, \mu, \tau$) with the mass eigenstates  
$|\nu_i\rangle$ ($i=1, 2, 3$): 
\be
|\nu_a\rangle = \sum_i U_{ai}^*\, |\nu_i\rangle \,.
\ee
In the 3f case the leptonic mixing matrix $U$ is a unitary $3\times 
3$ matrix, which in the standard parameterization can be written as
\begin{align}
U &=  O_{23} \Gamma_\delta O_{13} \Gamma^\dagger_\delta O_{12} \nonumber\\
\vspace*{1.5mm}
  &= \left( \begin{matrix} c_{12} c_{13} &
    s_{12} c_{13} & s_{13} {\rm e}^{-{\rm i} \delta_{\rm CP}} \\ -s_{12}
    c_{23} - c_{12} s_{13} s_{23} {\rm e}^{{\rm i} \delta_{\rm CP}} & c_{12}
   c_{23} - s_{12} s_{13} s_{23} {\rm e}^{{\rm i} \delta_{\rm CP}} & c_{13}
    s_{23} \\ s_{12} s_{23} - c_{12} s_{13} c_{23} {\rm e}^{{\rm i}
      \delta_{\rm CP}} & -c_{12} s_{23} - s_{12} s_{13} c_{23} {\rm
      e}^{{\rm i} \delta_{\rm CP}} & c_{13} c_{23} \end{matrix} \right) \,.
   \label{U1}
\end{align}
Here $O_{ij}$ is the orthogonal rotation matrix in the $ij$-plane which 
depends on the mixing angle $\theta_{ij}$, $\Gamma_\delta=\mathrm{diag} 
(1,1,{\rm e}^{{\rm i}\delta_{\rm CP}})$, $\delta_{\rm CP}$ being the  
Dirac-type CP-violating phase, $s_{ij} \equiv \sin \theta_{ij}$ and 
$c_{ij} \equiv \cos \theta_{ij}$. In the 3f case there are also, in general, 
two Majorana-type CP-violating phases; however, these phases 
do not affect neutrino oscillations, and I will not discuss them. 

Neutrino data allow two different neutrino mass orderings, 
the normal mass hierarchy and the inverted mass hierarchy (see 
Figs. \ref{scheme1} and \ref{scheme2}).  
%
\begin{samepage}
\begin{figure}[tbh] 
\centering
    \includegraphics[width=0.38\textwidth]{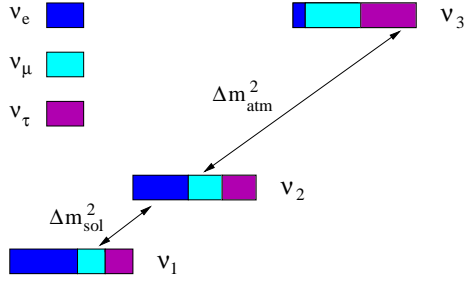}
\vspace*{2mm}
    \caption{\small Normal mass hierarchy. }   
\label{scheme1}
\vspace*{-2mm}
\end{figure}
%
\begin{figure}[h] 
\centering
    \includegraphics[width=0.38\textwidth]{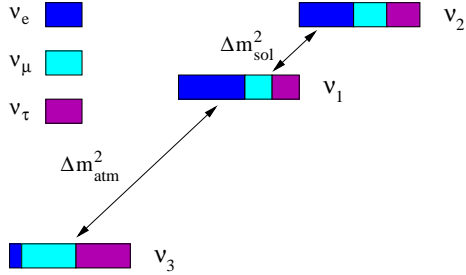}
\vspace*{2mm}
    \caption{\small Inverted mass hierarchy. }   
\label{scheme2}
\vspace*{-2mm}
\end{figure}
\end{samepage}
%
The lines of the matrix $U$ in Eq. (\ref{U1}) represent the neutrino 
flavour eigenstates in terms of mass eigenstates, whereas its columns give 
the mass eigenstates in terms of flavour eigenstates. In particular, the 
value of $|U_{e3}|^2$ is the weight of $\nu_e$ in the third mass eigenstate.

\section{ Three-flavour neutrino oscillations in matter}

Neutrino oscillations in matter are described by the Schr\"odinger-like 
evolution equation
\be
i\frac{d}{dt}
\left(\begin{array}{ccc}\!\! \nu_e\!\! \\ \!\! \nu_\mu\!\! \\
\!\! \nu_\tau \!\!\end{array}\right) \!
=\!\!\left[U \! \left( \begin{array}{ccc} \!\! E_1 \! & \! 0 \! 
& \! 0 \! \\ \!\! 0 \! & \! E_2 \! & \! 0 \! \\ \!\! 0 \! & \! 0 
\! & \! E_3 \! \end{array} 
\right) \! 
U^\dagger
+ \left( \begin{array}{ccc}\!\! V(t) \!\! & \! 0 & 0 \! \\ \!\! 0 \!\! 
& \! 0 & 0 \! \\ \!\! 0 \!\! & \! 0 & 0 \!
\end{array} \right) \! \right]\!\!
\left(\begin{array}{ccc}\!\! \nu_e \!\! \\ \!\! \nu_\mu \!\! \\ 
\!\nu_\tau  \!\!
\end{array}\right)\,. 
\label{Sc1}
\ee
Here $E_i$ are the neutrino eigenenergies in vacuum, and 
the effective potential $V=\sqrt{2}G_F N_e$ is due to the charged-current 
interaction of $\nu_e$ with the electrons of the medium \cite{MSW}.  
The neutral current induced potentials are omitted from Eq. (\ref{Sc1}) 
because they are the same for neutrinos of all three species and therefore 
do not affect neutrino oscillations. This, however, is only true in 
leading (tree) order; radiative corrections induce tiny differences 
between the neutral current potentials 
of $\nu_e$, $\nu_\mu$ and $\nu_\tau$ and, 
in particular, result in a very small $\nu_\mu$ -- $\nu_\tau$ 
potential difference $V_{\mu\tau}\sim 10^{-5}\, V$ \cite{BLM}. 
This quantity is negligible in most situations but may be important for 
supernova neutrinos.

For matter of constant density, closed-form solutions of the evolution 
equation can be found \cite{exact}; however, the corresponding expressions 
are rather complicated and not easily tractable. For a general electron 
density profile $N_e \ne const$ no closed-form solutions exist. It is
therefore desirable to have approximate analytic solutions of the neutrino 
evolution equation. A number of such solutions were found, most of them
based on the expansions in one (or both) of the two small parameters: 
$\Delta m_{21}^2/\Delta m_{31}^2 = \Delta m_{\rm sol}^2/\Delta m_{\rm atm}^2 
\simeq 0.03$, $|U_{e3}| = |\sin\theta_{13}| \aprle 0.2$ \cite{Concha}. For a 
recent discussion and a summary of the earlier results see ref. \cite{expan}.

In the limits $\Delta m_{21}^2=0\,$ or $\,U_{e3}=0$ the transition 
probabilities acquire an effective 2-flavour (2f) form. 
When both these parameters vanish, the genuine 2f case is recovered. 

\section{General properties of 3f oscillation probabilities}

In the 3f case, there are nine oscillation (survival and transition) 
probabilities for neutrinos and the same number for antineutrinos, hence 
altogether 18 probabilities. How many of them are independent? 

Consider first only neutrinos. Unitarity (probability conservation) 
gives 6 constraints
\be
\sum_b P_{ab}=\sum_a P_{ab}=1 \qquad (a, b= e,\mu,\tau)\,,
\label{unitar}
\ee
of which 5 are independent; this leaves 9-5=4 oscillation probabilities 
independent. However, recently it has been realized \cite{expan} that one 
can further reduce the number of independent oscillation probabilities. 
This is possible because the evolution equation (\ref{Sc1}) has a symmetry 
related to the fact that the matrix of matter-induced potentials $diag(V(t), 0, 
0)$ commutes with $O_{23}$. 
Let us define the ``$\theta_{23}$-transformed'' probabilities
\be
\tilde P_{ab}=P_{ab}(s_{23}^2\leftrightarrow c_{23}^2,\,
\sin 2\theta_{23}\to -\sin 2\theta_{23})
\ee
(this transformation can, e.g., be achieved through the shift $\theta_{23}\to 
\theta_{23}+\pi/2$). Then, by inspecting the properties of Eq. (\ref{Sc1}) with 
respect to the rotation by $O_{23}$, it is easy to show that
\be
P_{e\tau}=\tilde P_{e\mu}\,,\qquad
P_{\tau\mu}=\tilde P_{\mu\tau}\,,\qquad
P_{\tau\tau}=\tilde P_{\mu\mu}\,.
\ee
Two of these three relations are independent, which leaves us with 4-2=2 
independent probabilities. Not any two probabilities would do; one 
possible choice is $P_{e\mu}$ and $P_{\mu\tau}$.

Since the oscillation probabilities for antineutrinos $P_{\bar{a}\bar{b}}$ 
are related to those for neutrinos $P_{ab}$ through
\be
P_{\bar{a}\bar{b}}=P_{ab}(\delta_{\rm CP}\to -\delta_{\rm CP}, V\to -V)\,,
\ee
one concludes that all 18 neutrino and antineutrino probabilities can be 
expressed through just two \cite{expan}. 

Analogously, by rotating Eq. (\ref{Sc1}) with the matrix  
$O_{23}'=O_{23}\times diag(1, 1, e^{i\delta_{\rm CP}})$, one can study the 
general dependence of the oscillation probabilities on the CP-violating phase 
$\delta_{\rm CP}$ \cite{Yokomakura:2002av}. For example, for $P_{e\mu}$ and 
$P_{\mu\tau}$ one finds 
\bea
P_{e\mu} &=& A_{e\mu}\cos\delta_{\rm CP}+B_{e\mu}\sin\delta_{\rm CP}
+C_{e\mu}\,, \qquad\qquad \nonumber \\
P_{\mu\tau} &=& A_{\mu\tau}\cos\delta_{\rm CP}+B_{\mu\tau}
\sin\delta_{\rm CP}+C_{\mu\tau} 
{} +D_{\mu\tau}\cos 2\delta_{\rm CP}+E_{\mu\tau}\sin 2\delta_{\rm CP}\,.
\eea

\section{3f effects in neutrino oscillations}

\subsection{Two kinds of three-flavour effects }

There are two kinds 3f effects in neutrino oscillations. First, there are 
effects which, in a sense, are trivial. These include: 

\begin{itemize} 
\item The existence of new physical oscillation channels -- i.e., in 
addition to $\nu_e\leftrightarrow \nu_\mu$ there are $\nu_e\leftrightarrow
\nu_\tau$ and $\nu_\mu\leftrightarrow \nu_\tau$ channels; mutual influence
of the channels through unitarity;

\item New ``parameter channels'' for the same physical channel. For example, 
$\nu_e\leftrightarrow \nu_\mu$ oscillations can be governed by two pairs 
of parameters, $(\Delta m_{21}^2,\,\theta_{12})$ and $(\Delta m_{31}^2,\,
\theta_{13})$, corresponding to two ways in which these oscillations can 
occur. 

\end{itemize} 
Second, there are non-trivial effects, i.e. qualitatively new effects that  
are specific for three (or more) flavours and do not occur in the 
2f case:

\begin{itemize} 

\item  Fundamental CP- and T-violation;

\item Matter-induced T-violation; 

\item Interference of different ``parameter channels'' -- specific 
contributions to oscillation probabilities;

\item Matter effects on $\nu_\mu\leftrightarrow \nu_\tau$ oscillations. 

\end{itemize} 

\noindent
A characteristic feature of the non-trivial 3f effects (except for the last 
one) is that they disappear if at least one mixing angle is 0 or $90^\circ$, 
or at least one $\Delta m_{ij}^2=0$. 

I will discuss both types of 3f effects. 

\subsection{Matter effects in $\nu_\mu \leftrightarrow \nu_\tau$ 
oscillations} 

Since the matter-induced potentials for $\nu_\mu$ and $\nu_\tau$ are the same 
(neglecting the tiny radiative corrections), in the 2f case the $\nu_\mu 
\leftrightarrow \nu_\tau$ oscillations are not affected by matter. This, 
however, is not true in the 3f case; therefore the effect of matter on $\nu_\mu 
\leftrightarrow \nu_\tau$ oscillations is a pure 3f effect. It vanishes only 
when both $\Delta m_{21}^2$ and $U_{e3}$ vanish. The effects of the Earth's 
matter on $\nu_\mu \leftrightarrow \nu_\tau$ oscillations can manifest 
themselves in the long-baseline accelerator as well as in atmospheric neutrino 
experiments \cite{ADLS,nu2002}. It has been demonstrated recently that these 
effects can be rather large \cite{Gandhi:2004md}.

\subsection{Solar neutrinos }
In the 3f case, solar $\nu_e$ can in principle oscillate into either 
$\nu_\mu$, or $\nu_\tau$, or some their combination. 
What do they actually  oscillate to? 
 
It is easy to answer this question. 
The smallness of the mixing parameter $|U_{e3}|$ implies that the mass 
eigenstate $\nu_3$, separated by a large mass gap from the other two, 
is approximately given by 
\be
\nu_3 \simeq s_{23}\, \nu_\mu + c_{23}\,\nu_\tau 
\label{nu3}
\ee
and, to first approximation, does not participate in the solar neutrino 
oscillations. From the unitarity of the leptonic mixing matrix it then
follows that the solar neutrino oscillations are the oscillations between 
$\nu_e$ and a state $\nu'$ which is the linear combination of $\nu_\mu$ and 
$\nu_\tau$, orthogonal to $\nu_3$: 
\be
\nu' = c_{23}\,\nu_\mu - s_{23}\,\nu_\tau 
\label{nuprime}
\ee
Therefore for solar neutrinos 
\be
P(\nu_e\to \nu_\mu)/P(\nu_e\to \nu_\tau)\simeq c_{23}^2/s_{23}^2\,.
\label{nu4}
\ee
Since the mixing angle $\theta_{23}$, responsible for the atmospheric 
neutrino oscillations, is known to be close to $45^\circ$, 
Eq. (\ref{nu4}) implies that the solar $\nu_e$ oscillate into a 
superposition of $\nu_\mu$ and $\nu_\tau$ with equal or almost equal
weights. The same argument applies to the long-baseline oscillations of 
reactor antineutrinos (KamLAND experiment). For reactor experiments with 
relatively short baselines $L\simeq 1$ km (such as CHOOZ and Palo Verde), 
the same is true when $\theta_{13}\ll 0.03$. In the opposite limit,  
$\theta_{13} \gg 0.03$, one finds $P(\bar{\nu}_e\to \bar{\nu}_\mu)
/P(\bar{\nu}_e\to \bar{\nu}_\tau)\simeq s_{23}^2/c_{23}^2$, which is also 
close to unity. In the intermediate case, $\theta_{13}\sim 0.03$,  
deviations from unity are possible due to the interference terms in the 
probabilities $P(\bar{\nu}_e\to \bar{\nu}_\mu)$ and $P(\bar{\nu}_e\to 
\bar{\nu}_\tau)$. 

What are the 3f effects in the oscillation probabilities of solar 
neutrinos?  Since at low energies $\nu_\mu$ and $\nu_\tau$ are experimentally 
indistinguishable, all the observables depend on just one probability -- the 
$\nu_e $ survival probability $P(\nu_e\to \nu_e)$. The loss of coherence of  
the neutrino state in the course of neutrino propagation between the Sun and 
the Earth leads to an effective averaging over fast oscillations due to 
the large mass squared difference $\Delta m_{\rm atm}^2=\Delta 
m_{31}^2$, which yields \cite{Lim}
\be
P(\nu_e\to \nu_e) \simeq c_{13}^4 \tilde{P}_{2ee}(\Delta m_{21}^2, 
\theta_{12}, N_{\rm eff}) + s_{13}^4\,.
\label{Psol}
\ee
Here $\tilde{P}_{2ee}(\Delta m_{21}^2,\theta_{12}, N_{\rm eff})$ is the 2f
survival probability of $\nu_e$ in matter with the effective 
electron density $N_{\rm eff}=c_{13}^2\,N_e$. 

As follows from the CHOOZ data \cite{Chooz}, the second term 
in Eq. (\ref{Psol}), $s_{13}^4$,  does not exceed $10^{-3}$, i.e. is 
negligible. 
At the same time, the coefficient $c_{13}^4$ of $\tilde{P}_{2ee}$ in the 
first term may differ from unity by as much as $\sim 5$ -- 10\,\%. Thus, 
3f effects may lead to an approximately energy-independent suppression of 
the $\nu_e$ survival probability by up to 10\%.  With high precision solar 
data this must be taken into account.  

{}From Eq.~(\ref{Psol}) it is clear that the fluxes of various components 
of the solar neutrino spectrum $f_i$ ($i=pp$, $^7$Be, $^8$B, ...) are always 
extracted from the charged-current experimental data in the combinations 
$f_i c_{13}^4$. This leads to an intrinsic uncertainty in their values due to 
the uncertainty in $\theta_{13}$. 
In contrast to this, the neutral currents experiments give the fluxes which 
are free of {\em both} astrophysics and $\theta_{13}$ uncertainties. 

\vspace*{-3mm}
\begin{figure}[h]
\epsfxsize=16.0cm\epsfbox{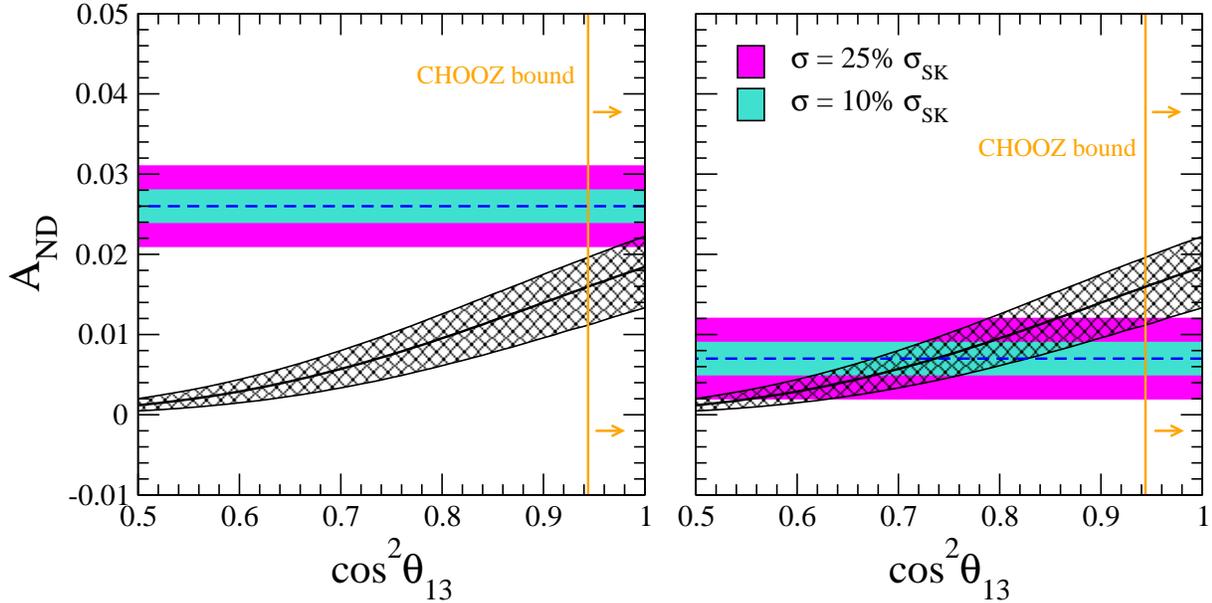}
\vspace*{2mm}
    \caption{\small Expected night-day asymmetry at UNO and Hyper-Kamiokande 
(horizontal bands) with central values larger (left panel) and smaller (right 
panel) than the current SK one. Hatched areas are theoretical expectations 
based on 3$\sigma$ allowed regions of $\Delta m_{21}^2$ and $\theta_{12}$ 
from KamLAND and solar neutrino data. Regions 
allowed by CHOOZ (3$\sigma$) are to the right of vertical lines. 
} 
\label{fig:Adn}
\vspace*{-2mm}
\end{figure}
While the day-time survival probability of solar $\nu_e$ (\ref{Psol}) scales 
essentially as $c_{13}^4$, the day-night signal difference due to the Earth 
matter effect scales as $c_{13}^6$ \cite{Blennow:2003xw,Akhmedov:2004rq}. 
Unfortunately, the experimental errors of the day-night asymmetry measured 
by the Super-Kamiokande (SK) and SNO detectors are too large and no useful 
information on $\theta_{13}$ can presently be extracted (the SK value, which 
has a smaller error, is $A_{ND}=2.1 \% \pm 2.0\% (\rm{stat}) \pm 1.3 \% 
(\rm{syst})$). 
However, future large water Cherenkov detectors, such as UNO or 
Hyper-Kamiokande, may be able to probe $\theta_{13}$ \cite{Akhmedov:2004rq}. 
This is illustrated in Fig. \ref{fig:Adn}, which shows the night-day signal 
asymmetry $A_{ND}$ in future detectors. 
Fig. \ref{fig:Adn} allows for possible deviations of the central value of the 
observed asymmetry from the current SK one within $1\sigma$ error of the latter. 
As one can see from the left panel, if the future central value of measured
$A_{ND}$ is higher than the present SK one, the current upper limit on 
$\theta_{13}$ can be improved. If, on the contrary, a lower value of 
$A_{ND}$ is measured (right panel), the derived upper limit on $\theta_{13}$
will be substantially weaker than the current one and thus irrelevant. However, 
as can be seen from the figure, 
in that case a {\em lower} bound on $\theta_{13}$ may appear; 
together with the current upper bound it may actually lead to a rather 
precise determination of $\theta_{13}$ \cite{Akhmedov:2004rq}. 
If the future central value of $A_{ND}$ coincides with the current SK one, 
no useful information on $\theta_{13}$ can be obtained.

\subsection{Atmospheric neutrinos }

\hspace{0.6cm}
(1) The dominant channel $\nu_\mu \leftrightarrow \nu_\tau$. 
In the 2f limit, there are no matter effects in this channel (neglecting 
tiny $V_{\mu\tau}$ caused by radiative corrections). The oscillation 
probability is independent from the sign of $\Delta m_{31}^2$, 
i.e. cannot differentiate between the normal and inverted neutrino mass 
hierarchies. The 3f effects result in a sensitivity to matter effects and to 
the sign of $\Delta m_{31}^2$.  

(2) The subdominant channels $\nu_e \leftrightarrow \nu_{\mu,\tau}$. 
Contributions of these oscillation channels to the number of $\mu$ -- like 
events are subleading and difficult to observe.  For e-like events, one could 
{\it a priori} expect significant oscillations effects. 
However, these effects are in fact strongly suppressed because of the 
specific composition of the atmospheric neutrino flux and proximity of the 
mixing angle $\theta_{23}$ to $45^\circ$.  
Indeed, in the 2f limits one finds 
\be
~~\frac{F_e-F_e^0}{F_e^0} = \tilde{P}_2(\Delta m_{31}^2,\,\theta_{13},
V)\cdot (r s_{23}^2-1)
\label{relat1}
\ee
in the limit $\Delta m_{21}^2\to 0$ \cite{ADLS}, and  
\be
~~\frac{F_e-F_e^0}{F_e^0} = \tilde{P}_2(\Delta m_{21}^2,\,\theta_{12},
V)\cdot (r c_{23}^2-1)
\label{relat2}
\ee
in the limit $s_{13}\to 0$ \cite{PS1999}. 
Here $F_e^0$ and $F_e$ are the $\nu_e$ fluxes in the absence and in the 
presence of the oscillations, respectively, and $r \equiv F_\mu^0/F_e^0$.  
At low energies $r \simeq 2$; also, we know that $s_{23}^2\simeq c_{23}^2 
\simeq 1/2$. Therefore the factors $(r s_{23}^2-1)$ and $(r c_{23}^2-1)$ 
in Eqs. (\ref{relat1}) and (\ref{relat2}) are very small and strongly
suppress the oscillation effects even if the transition probabilities 
$\tilde{P}_2$ are close to unity. This happens because of the strong 
cancellations of the transitions from and to the $\nu_e$ state. 

All this looks as a conspiracy to hide the oscillation effects on the e-like 
events! This conspiracy is, however, broken by the 3f effects. Keeping both 
$\Delta m_{21}^2$ and $s_{13}$ in leading order yields \cite{Peres:2003wd}. 
\bea
\frac{F_e-F_e^0}{F_e^0} ~\simeq~ \tilde{P}_2(\Delta 
m_{31}^2,\,\theta_{13}) \cdot (r\,s_{23}^2-1) 
~+~ \tilde{P}_2(\Delta m_{21}^2,\,\theta_{12}) \cdot (r\,c_{23}^2-1) 
\nonumber \\
~-~ 2 s_{13}\, s_{23}\, c_{23}\, r\, {\rm
Re} (\tilde{A}_{ee}^* \, \tilde{A}_{\mu e})\,.~
\label{relat3}
\eea
Here $\tilde{A}_{ee}$ and $\tilde{A}_{\mu e}$ are the $\nu_e$ survival and 
transition amplitudes in the rotated basis $\tilde{\nu}\approx 
O_{13}^\dagger O_{23}^\dagger \nu_{fl}$, where 
$\nu_{fl}$ is the neutrino state in the flavour basis. The interference term, 
which represents the genuinely 3f effects, is not suppressed by the flavour 
composition of the atmospheric neutrino flux; it can reach a few per cent and 
may be partially responsible for some excess of the upward-going sub-GeV e-like 
events observed at Super-Kamiokande. 
However, this term seems to be insufficient to fully explain the excess, 
which may be a hint of a deviation of $\theta_{23}$ from $ 45^\circ$ 
\cite{Peres:2003wd}.

\subsection{Reactor antineutrinos}
For reactor neutrino experiments, the $\bar{\nu}_e$ survival probability can 
be written as 
\be
\hspace*{-1.5cm}
P_{\bar{e}\bar{e}} \simeq  1-\sin^2 2\theta_{13} \cdot \sin^2\left(
\frac{\Delta m_{31}^2}{4E} L\right)-c_{13}^4 \sin^2 2\theta_{12} \cdot
\sin^2\left(\frac{\Delta m_{21}^2}{4E}L\right)\,.
\label{Preac1}
\ee
Since the average energy of reactor $\bar{\nu}_e$'s is $\bar{E} \sim 4$ 
MeV, for intermediate-baseline experiments, such as CHOOZ and Palo 
Verde ($L\aprle 1$ km), one has 
\be 
\frac{\Delta m_{31}^2}{4E}\,L\sim 1\,, \quad\quad 
\frac{\Delta m_{21}^2}{4E}\,L \ll 1 \,.
\label{cond1}
\ee
This justifies the use of the one mass scale dominance approximation, 
in which the last term in (\ref{Preac1}) is neglected:
\be
P(\bar{\nu}_e\to\bar{\nu}_e) = 1- \sin^2 2\theta_{13} \cdot
\sin^2\left(\!\frac{\Delta m_{31}^2}{4E}L\!\right)\,.
\label{Preac2}
\ee
This is a pure 2f result. Note, however, that disregarding the last term in 
(\ref{Preac1}) is only legitimate if $\theta_{13}$ is larger than  $\sim 0.03$, 
which is about the reach of the currently discussed next-generation reactor 
neutrino experiments.

For KamLAND, which is a very long baseline reactor experiment 
($\bar{L}\simeq 170$ km), one has 
\be
\frac{\Delta m_{31}^2}{4E}\,L\gg 1\,, \qquad\quad 
\frac{\Delta m_{21}^2}{4E}\,L \aprge 1 \,. 
\label{cond2}
\ee
Averaging over the fast oscillations driven by 
$\Delta m_{31}^2=\Delta m_{\rm atm}^2 $ yields  
\be
P(\bar{\nu}_e\to \bar{\nu}_e) = c_{13}^4 P_{2 \bar{e} \bar{e}}(\Delta
m_{21}^2, \theta_{12}) + s_{13}^4\,.
\label{Preac3}
\ee
This has the same form as Eq. (\ref{Psol}).
The 2f survival probability $P_{2 \bar{e} \bar{e}}$ is, in first approximation, 
just the corresponding $\bar{\nu}_e$ survival probability in vacuum, which 
can be obtained from (\ref{Preac2}) by substituting $\theta_{13}\to 
\theta_{12}$, $\Delta m_{31}^2\to \Delta m_{21}^2$. Note, however, that matter 
effects in KamLAND can reach a few per cent, i.e. can be comparable with the 
effects of non-zero $\theta_{13}$, and so should be taken into account in 3f 
analyses. The probability (\ref{Preac3}) can differ from the 2f probability 
$P_{2 \bar{e} \bar{e}}$ by up to $\sim 10\%$.

\subsection{Long-baseline accelerator experiments}

(1) $\nu_\mu$ disappearance. 

3f effects can result in up to $\sim 10$\% corrections to the disappearance 
probability, mainly due to the factor $c_{13}^4$ in the effective 
amplitude of the $\nu_\mu\leftrightarrow \nu_\tau$ oscillations, 
%
%
$\sin^2 (2\theta_{\mu\tau})_{\rm eff} \equiv c_{13}^4\,\sin^2 2\theta_{23}$.
Another manifestation of 3-flavourness are matter effects in 
$\nu_\mu\leftrightarrow \nu_\tau$ oscillations. The same applies to 
$\nu_\tau$ appearance in experiments with the conventional neutrino beams. 
$\nu_\mu$ disappearance receives contributions also from the 
subdominant $\nu_\mu\leftrightarrow \nu_e$ oscillations. 

(2) $\nu_\mu$ appearance at neutrino factories; $\nu_e$ appearance at 
neutrino factories and in experiments with the conventional neutrino beams. 

These are driven by the $\nu_e \leftrightarrow \nu_{\mu,\tau}$ oscillations. 
There are two channels through which these subdominant oscillations can 
proceed -- those governed by the parameters $(\theta_{13},~\Delta m_{31}^2)$ 
and $(\theta_{12},~\Delta m_{21}^2)$. For typical energies of the long-baseline 
(LBL) accelerator experiments, a few GeV to tens of GeV, one finds that for 
$\theta_{13}$ in the range $3\cdot 10^{-3}\,\aprle\, \theta_{13}\, \aprle\, 
3\cdot 10^{-2}$ the two channels compete; otherwise one of them dominates.  

Unlike in the case of atmospheric neutrinos, there is no suppression of the 
oscillation effects on the $\nu_e$ flux due to the flavour composition 
of the original flux.

The dependence of the oscillation probabilities on the CP-violating phase 
$\delta_{\rm CP}$ comes from the interference terms and is a pure 
3f effect. The 3f effects will be especially important for the future 
experiments at neutrino factories which are designed for precision 
measurements of neutrino parameters.

\subsection{Supernova neutrinos}
In supernovae, matter density varies in a very wide range, and the 
conditions for three MSW \cite{MSW} resonances are satisfied (taking into 
account that due to radiative corrections $V_{\mu\tau}\ne 0$).
The hierarchy $\Delta m^2_{21}\ll \Delta m^2_{31}$ leads to the approximate 
factorization of the transition dynamics at the resonances, so that the 
transitions, to first approximation, are effectively 2f ones. However, the 
observable effects of the supernova neutrino oscillations depend 
on the transitions between all three neutrino species \cite{DS}. 

Supernova neutrinos can propagate significant distances inside the Earth  
before reaching the detector. Matter effects on the oscillations 
of supernova neutrinos inside the Earth depend crucially on the sequence 
of the neutrino flavour conversions in the supernova which, in turn, depends 
on the sign of $\Delta m_{31}^2$ and is very sensitive to the value of the 
leptonic mixing parameter $U_{e3}$. Thus, the Earth matter effects on 
supernova neutrinos can be used to determine the sign of $\Delta m_{31}^2$ 
and to probe $|U_{e3}|$ down to very small values ($\sim 10^{-3}$) \cite{SN1}. 
The transitions due to the $\nu_\mu - \nu_\tau$ potential difference 
$V_{\mu\tau}$ caused by radiative corrections may have observable consequences 
if the originally produced $\nu_\mu$ and $\nu_\tau$ fluxes are not exactly 
the same \cite{SN2}.

If neutrinos are Majorana particles, a combination of the MSW effect and 
resonance spin-flavour precession (RSFP) due to the interaction of neutrino 
transition magnetic moments $\mu_{\nu}$ with supernova magnetic fields $B$ 
can result in the conversion $\nu_e\to\bar{\nu}_e$. Such a conversion would lead 
to the transformation of the supernova $\nu_e$, born in the neutronization 
process, into their antiparticles. This effect would have a clear experimental 
signature and its observation would be a smoking gun evidence for the neutrino 
transition magnetic moments. It would also signify the leptonic mixing parameter 
$|U_{e3}|$ in excess of $10^{-2}$.
The conversion mechanism is efficient if $\mu_\nu B_{res}\aprge 10^{-13}
\mu_B\cdot 10^9$ G. In the 2f approach, the  $\nu_e\to\bar{\nu}_e$ transition is 
only possible in the case of the inverted neutrino mass hierarchy 
\cite{Ando:2003is}.
However, in the full 3f framework one finds a new RSFP resonance to exist,  due 
to which the $\nu_e\to\bar{\nu}_e$ conversion can occur also for the normal mass 
hierarchy \cite{Akhmedov:2003fu}. 
Thus, the possibility of $\nu_e\to\bar{\nu}_e$ transitions of supernova 
neutrinos in the case of the normal neutrino mass hierarchy is a pure 3f 
effect.

\section{CP and T violation in $\nu$ oscillations in vacuum}

The probability of $\nu_a \to \nu_b$ oscillations in vacuum is given by
\be
P(\nu_a,t_0 \to\nu_b; t)=\left|\sum_i U_{bi} e^{-i E_i (t-t_0)}
U_{ai}^*\right|^2\!. 
\label{vac}
\ee
In the general case of $n$ flavours the leptonic mixing matrix $U_{ai}$ depends 
on $(n-1)(n-2)/2$ Dirac-type CP-violating phases $\{\delta_{\rm CP}\}$. 

Under CP transformation, neutrinos are replaced by their antiparticles
($\nu_{a,b}\leftrightarrow \bar{\nu}_{a,b}$), which is equivalent to 
the complex conjugation of $U_{ai}$: 
\bea
{\rm CP:}\quad 
\qquad\qquad
\nu_{a,b}\leftrightarrow \bar{\nu}_{a,b}\qquad\qquad
\qquad\qquad\qquad\qquad
\nonumber \\
\Leftrightarrow~ U_{ai} \to U_{ai}^* \quad(\{\delta_{\rm CP}\} \to
-\{\delta_{\rm CP}\})\,. ~~~
\label{CPvac}
\eea

Time reversal transformation interchanges the initial and final evolution 
times $t_0$ and $t$ in Eq. (\ref{vac}),  i.e. corresponds to evolution 
``backwards in time''. As follows from Eq. (\ref{vac}), the interchange 
$t_0 \buildrel \rightarrow \over {_{\leftarrow}} t$ is equivalent to the 
complex conjugation of the exponential factors in the oscillation 
amplitude. Since the transition probability only depends on the modulus of 
the amplitude, this is equivalent to the complex conjugation of the factors 
$U_{bi}$ and $U_{ai}^*$, which in turn amounts to interchanging 
$a\buildrel \rightarrow \over {_{\leftarrow}} b$. Thus, instead of
evolution ``backwards in time'' one can consider evolution forward in
time, but between the interchanged initial and final flavours: 
\bea
T:  
\qquad\qquad
~~t_0\buildrel \rightarrow \over {_{\leftarrow}} t 
\Leftrightarrow \nu_{a} \leftrightarrow \nu_{b}  
\qquad\qquad\qquad\qquad\qquad
\nonumber \\
\Rightarrow  ~U_{ai} \to U_{ai}^*
~~(\{\delta_{\rm CP}\} \to - \{\delta_{\rm CP}\})\,. 
\qquad
\label{Tvac}
\eea

Under the combined action of CP and T one has 
\bea
{\rm CPT}: \quad 
\qquad\qquad
\nu_{a,b} \leftrightarrow \bar{\nu}_{a,b} ~~{\rm and}~~ 
t_0\buildrel \rightarrow \over {_{\leftarrow}} t 
~(\nu_a \leftrightarrow \nu_b) 
\nonumber \\ 
\quad\quad\quad\quad\quad \Rightarrow ~P(\nu_a\to\nu_b) \to P(\bar{\nu}_b
\to \bar{\nu}_a)\,.~
\label{CPT}
\eea
{}From CPT invariance it follows that CP violation
implies T violation and vice versa. 

CP and T violation can be characterized by the probability differences 
\be
\Delta P_{ab}^{\rm CP}\equiv P(\nu_a\to \nu_b)-P(\bar{\nu}_a\to\bar{\nu}_b)\,,
\label{DCP}
\ee
\be
\Delta P_{ab}^{\rm T}\, \equiv\, P(\nu_a\to \nu_b)-P(\nu_b\to\nu_a)\,.
\label{DT}
\ee
{}From CPT invariance it follows that the CP- and T-violating probability 
differences coincide, and that the survival probabilities 
have no CP asymmetry: 
\be
\Delta P_{ab}^{\rm CP} = \Delta P_{ab}^{\rm T}\,; 
\quad\quad \Delta P_{aa}^{\rm CP} = 0\,.
\label{CPT2}
\ee
CP and T violations are absent in the 2f case, so any observable violation
of these symmetries in neutrino oscillations in vacuum would be a pure
$\ge 3$f effect. 

In the 3f case, there is only one CP-violating Dirac-type phase
$\delta_{\rm CP}$ and so only one CP-odd (and T-odd) probability
difference: 
\be
\Delta P_{e\mu}^{\rm CP} ~=~ \Delta P_{\mu\tau}^{\rm CP} ~=~ 
\Delta P_{\tau e}^{\rm CP} ~\equiv~ \Delta P\,,
\label{DeltaP}
\ee
where
\be
\Delta P = {} -\, 4 s_{12}\,c_{12}\,s_{13}\,c_{13}^2\,s_{23}\,c_{23}\,
\sin\delta_{\rm CP}\,
\left[\sin\left(\!\frac{\Delta m_{12}^2}{2E} L\!
\right)\!+\sin\left(\!\frac{\Delta m_{23}^2}{2E} L\!\right) \!+ \sin\left(
\!\frac{\Delta m_{31}^2}{2E} L\!\right)\! \right].
\ee
It vanishes 

$\bullet$ when at least one $\Delta m_{ij}^2 = 0$

$\bullet$ when at least one $\theta_{ij} = 0$ or $90^\circ$ 

$\bullet$ when $\delta_{\rm CP} = 0$ or $180^\circ$

$\bullet$ in the averaging regime 

$\bullet$ in the limit $L\to 0$ (as $L^3$) \\
Clearly, this quantity is very difficult to observe.  

\section{CP- and T-odd effects in $\nu$ oscillations in matter}

For neutrino oscillations in matter, CP transformation (substitution 
$\nu_a\leftrightarrow \bar{\nu}_a$) implies not only complex conjugating 
the leptonic mixing matrix, but also flipping the sign of the 
matter-induced neutrino potentials: 
\bea
{\rm CP:} \quad 
\qquad\qquad
U_{ai} \to U_{ai}^* ~(\{\delta_{\rm CP}\} \to - \{\delta_{\rm CP}\})\,,
\nonumber \\
V(r) \to {}- V(r)\,. \qquad\quad\,  
\label{CPmat}
\eea

It can be shown \cite{AHLO} that in matter with an arbitrary density profile, 
just as well as in vacuum, the action of time reversal on neutrino oscillations
is equivalent to interchanging the initial and final neutrino flavours. It 
is also equivalent to complex conjugating $U_{ai}$ and replacing the matter 
density profile by the reverse one: 
\bea
{\rm T:} \quad
\qquad\qquad
U_{ai} \to U_{ai}^* ~(\{\delta_{\rm CP}\} \to - \{\delta_{\rm CP}\})\,,
\nonumber \\
V(r) \to \tilde{V}(r)\,.  ~\quad\quad\quad\quad
\label{Tmat}
\eea
Here 
\be
\tilde{V}(r) = \sqrt{2} G_F \tilde{N} (r)\,,
\label{Vrev}
\ee
$\tilde{N} (r)$ being the reverse profile, i.e. the profile that corresponds 
to the interchanged positions of the neutrino source and detector. In the 
case of symmetric matter density profiles (e.g., matter of constant density), 
$\tilde{N} (r) = N (r)$. 

An important point is that the very presence of matter (with unequal numbers
of particles and antiparticles) violates C, CP and CPT, leading to CP-odd 
effects in neutrino oscillations even in the absence of the fundamental 
CP-violating phases $\{\delta_{\rm CP}\}$. This fake (extrinsic) CP violation 
may complicate the study of the fundamental (intrinsic) one.

\subsection{CP-odd effects in matter}

Unlike in vacuum, CP-odd effects in neutrino oscillations in matter exist 
even in the 2f case (in the case of three or more flavours, even when all
$\{\delta_{\rm CP}\}=0$):  
\be 
P(\nu_a\to \nu_b) \ne P(\bar{\nu}_a\to \bar{\nu}_b) \,.
\label{diff}
\ee
This is actually a well known fact -- for example, the MSW effect can
enhance the $\nu_e\leftrightarrow \nu_\mu$ oscillations and suppress the 
$\bar{\nu}_e\leftrightarrow \bar{\nu}_\mu$ ones or vice versa.
Moreover, in matter the survival probabilities are not CP-invariant:
\be
P(\nu_a\to \nu_a) \ne P(\bar{\nu}_a\to \bar{\nu}_a) \,.
\label{CPnoninv}
\ee
To disentangle fundamental CP violation from the matter induced one in
the LBL experiments one would need to measure the energy dependence of the 
oscillated signal or the signals at two baselines, which is a difficult 
task. The (difficult) alternatives are: 

$\bullet$ 
LBL experiments at relatively low energies and moderate baselines ($E\sim$
0.1 -- 1 GeV, $L\sim$ 100 -- 1000 km) \cite{lowE} -- in this case
matter effects are negligible.  

$\bullet$ Indirect measurements through \\ 
$~~~~~~\,$ (A) CP-even terms $\sim \cos\delta_{\rm CP}$ \cite{CPeven};  

\noindent
$~~~~~~\!$ (B) Area of leptonic unitarity triangle \cite{triangle}. \\
CP-odd effects cannot be studied in the supernova neutrino experiments
because of the experimental indistinguishability of low-energy $\nu_\mu$
and $\nu_\tau$.

\subsection{T-odd effects in matter}

Since CPT is not conserved in matter,  CP and T violations are no longer 
directly connected (although some relations between them still exist 
\cite{AHLO,MNP}). 
Therefore T-odd effects in neutrino oscillation in matter deserve 
an independent study. Their characteristic features are:  

$\bullet$ Matter does not necessarily induce T-odd effects (only
asymmetric matter with $\tilde{N}(r) \ne N(r)$ does).  

$\bullet$ There is no T violation (either fundamental or matter 
induced) in the 2f case. This is a simple consequence of unitarity.
For example, for the $(\nu_e, \nu_\mu)$ system one has 
\be 
P_{ee} + P_{e\mu} =1\,, \qquad\quad
P_{ee} + P_{\mu e} =1\,,
\label{unit1}
\ee
from which $P_{e\mu} = P_{\mu e}$. 

$\bullet$ In the 3f case there is only one T-odd probability difference
for $\nu$'s (and one for $\bar{\nu}$'s), irrespective of the matter
density profile:  
\be 
\Delta P_{e\mu}^T=\Delta P_{\mu\tau}^T=\Delta P_{\tau e}^T \,.
\label{unit2}
\ee
This is a consequence of 3f unitarity \cite{KP1988}. 

The matter-induced T-odd effects are very interesting, pure $\ge$3f
matter effects, absent in symmetric matter (in particular, in 
constant-density matter). They do not vanish in the regime 
of complete averaging of neutrino oscillations \cite{AHLO}. They may fake
the fundamental T violation and complicate its study, i.e. the extraction
of $\delta_{\rm CP}$ from the experiment. The matter-induced T-violating  
effects vanish when either $U_{e3}=0$ or $\Delta m_{21}^2=0$ (i.e., in the 2f 
limits) and so are doubly suppressed by both these small parameters. This 
implies that the perturbation theory can be used to obtain analytic 
expressions for the T-odd probability differences \cite{AHLO}.  

In an asymmetric matter, both fundamental and matter-induced T violations 
contribute to the T-odd probability differences $\Delta P_{ab}^T$. This 
may hinder the experimental determination of the fundamental CP- and
T-violating phase $\delta_{\rm CP}$. In particular, in the accelerator 
LBL experiments one has to take into account that the Earth's density 
profile is not perfectly spherically symmetric. To extract the fundamental 
T violation, strictly speaking one would need to measure 
\be
P_{\rm dir}(\nu_a\to \nu_b) -
P_{\rm rev}(\nu_b\to \nu_a)\,, 
\label{deltaP}
\ee
where $P_{\rm dir}$ and $P_{\rm rev}$ correspond to the direct and reverse matter 
density profiles. (An interesting point is that even the survival probabilities 
$P_{\mu\mu}$ and $P_{\tau\tau}$ can be used for that \cite{FK}). 

In practical terms, it would certainly be difficult to measure the 
quantity in (\ref{deltaP}): It would not be easy, for example, to move 
CERN to Gran Sasso and the Gran Sasso Laboratory to CERN. Fortunately, this 
is not actually necessary -- matter-induced T-odd effects due to imperfect 
sphericity of the Earth's density distribution are very small. They cannot 
spoil the determination  of $\delta_{\rm CP}$ if the error in 
$\delta_{\rm CP}$ is $>1$\% at 99\% C.L. \cite{AHLO}.  


\subsection{``CPT in matter''}

As was pointed out before, a matter with unequal numbers of particles and 
antiparticles violates CPT. Is there any relation between CP and T violations in 
matter which can play a role similar to the CPT relation in vacuum? For symmetric 
density profiles ($\tilde{V}(r)=V(r)$) such a relation was found in 
\cite{MNP}:
\be
P(\nu_a\to \nu_b; \,\delta_{\rm CP}, V(r)) ~=~
P(\bar{\nu}_b\to \bar{\nu}_a;\, \delta_{\rm CP}, -V(r))\,.
\ee
It is easy to generalize this to the case of an arbitrary density profile:
\be
P(\nu_a\to \nu_b;\, \delta_{\rm CP}, V(r)) ~=~
P(\bar{\nu}_b\to \bar{\nu}_a;\, \delta_{\rm CP}, -\tilde{V}(r))\,.
\ee
Unlike CPT in vacuum, this ``CPT in matter'' relation does not directly relate 
observables (there is no anti-Earth),  
and so is of limited practical use. However, it can be useful for 
cross-checking theoretical calculations of oscillation probabilities.

\section{Why study $U_{e3}$? (A hymn to $U_{e3}$)}

The leptonic mixing parameter $U_{e3}$ plays a very special role in
neutrino physics.  It is of particular interest for a number of reasons. 

First, it is the least known of leptonic mixing parameters: while we
have (relatively small) allowed ranges for the other two mixing parameters, 
we only know an upper bound on $|U_{e3}|$. 
Its smallness, which looks strange in the light of the fact that 
the other two mixing parameters, $\theta_{12}$ and $\theta_{23}$, are 
apparently large, remains essentially unexplained. (There are, however, some 
ideas which relate the smallness of $|U_{e3}|$ to that of $\Delta 
m_{\rm sol}^2/\Delta m_{\rm atm}^2$ 
\cite{ABR}). 

The smallness of $U_{e3}$ is likely to be the bottleneck for studying
the fundamental CP and T violation effects and matter-induced T-odd  
effects in neutrino oscillations. The same applies to the determination
of the sign of $\Delta m_{31}^2$ in future LBL experiments, which would 
allow us to discriminate between the normal and inverted neutrino mass 
hierarchies. Therefore it would be vitally important to know how small 
$U_{e3}$ actually is.  

The parameter $U_{e3}$ can be efficiently used to discriminate between 
various neutrino mass models \cite{Barr}. It is one of the main 
parameters that drives the subdominant oscillations of atmospheric 
neutrinos and is important for their study. 
It also governs the Earth matter effects on supernova neutrino 
oscillations. 

The parameter $U_{e3}$ drives the parametric amplification of oscillations of 
core-crossing neutrinos inside the Earth, which is an interesting matter effect, 
different from the MSW resonance enhancement \cite{SmTalk}.

And finally, $U_{e3}$ apparently provides us with the only opportunity to 
see the ``canonical'' MSW effect. While matter effects can be important 
even in the case of large vacuum mixing angles, the most spectacular
phenomenon, strong enhancement of mixing by matter, can only occur  
if the vacuum mixing angle is small. From what we know now, 
it seems that the only small leptonic mixing parameter is $U_{e3}$. 

All this makes measuring $U_{e3}$ one of the most important problems in 
neutrino physics.

\section{Conclusions}

3f effects in solar, atmospheric, reactor and supernova neutrino 
oscillations and in LBL accelerator neutrino experiments may be quite 
important. They can lead to up to $\sim 10$\% corrections to the 
oscillation probabilities and also to specific effects, absent in the 2f 
case. The manifestations of $\ge 3$ flavours in neutrino oscillations
include fundamental CP violation and T violation, matter-induced T-odd 
effects, matter effects in $\nu_\mu \leftrightarrow \nu_\tau$
oscillations, and specific CP- and T-conserving interference terms 
(proportional to the sines of three different mixing angles) in 
oscillation probabilities. The leptonic mixing parameter $U_{e3}$ plays a 
very special role and its study is of great interest.

\end{document}